\documentclass{article} 

\usepackage{preprint}
\usepackage{comment}
\usepackage{amsmath, amsthm, amssymb, amsfonts}

\usepackage{graphicx}
\usepackage{xcolor}
\usepackage{comment}
\usepackage{makecell}
\usepackage{amsmath}
\usepackage{cite}
\usepackage{booktabs}
\usepackage{makecell}
\usepackage{caption}
\usepackage{pdflscape}
\usepackage{float}

\usepackage[numbers,sort&compress]{natbib}
\usepackage[utf8]{inputenc}	
\usepackage[T1]{fontenc}	
\usepackage{xcolor}		
\usepackage[colorlinks = true,
            linkcolor = purple,
            urlcolor  = blue,
            citecolor = cyan,
            anchorcolor = black]{hyperref}	
\usepackage{booktabs} 		
\usepackage{nicefrac}		
\usepackage{microtype}		
\usepackage{lineno}		
\usepackage{float}			

\usepackage{lipsum}		
\usepackage{amsmath}
\usepackage{newfloat}
\DeclareFloatingEnvironment[name={Supplementary Figure}]{suppfigure}
\usepackage{sidecap}
\sidecaptionvpos{figure}{c}
\usepackage{ifthen}
\usepackage{xspace}
\newboolean{uprightparticles}
\setboolean{uprightparticles}{false}

\def\MagUp {\mbox{\em Mag\kern -0.05em Up}\xspace}

\ifthenelse{\boolean{uprightparticles}}%
{

 \def\PDelta      {\ensuremath{\Delta}\xspace}                 
 \def\PXi      {\ensuremath{\Xi}\xspace}                 
 \def\PLambda      {\ensuremath{\Lambda}\xspace}                 
 \def\PSigma      {\ensuremath{\Sigma}\xspace}                 
 \def\POmega      {\ensuremath{\Omega}\xspace}                 
 \def\PUpsilon      {\ensuremath{\Upsilon}\xspace}                 
 
 \def\PB      {\ensuremath{\mathrm{B}}\xspace}                 
                  
 \def\PD      {\ensuremath{\mathrm{D}}\xspace}

 \def\PK      {\ensuremath{\mathrm{K}}\xspace}

 \def\Pi      {\ensuremath{\mathrm{i}}\xspace}

}
{

 \mathchardef\PDelta="7101
 \mathchardef\PXi="7104
 \mathchardef\PLambda="7103
 \mathchardef\PSigma="7106
 \mathchardef\POmega="710A
 \mathchardef\PUpsilon="7107
                  
 \def\PB      {\ensuremath{B}\xspace}                 
                  
 \def\PD      {\ensuremath{D}\xspace}

 \def\PK      {\ensuremath{K}\xspace}

 \def\Pi      {\ensuremath{i}\xspace}

}

\makeatletter
\ifcase \@ptsize \relax
  \newcommand{\miniscule}{\@setfontsize\miniscule{4}{5}}
\or
  \newcommand{\miniscule}{\@setfontsize\miniscule{5}{6}}
\or
  \newcommand{\miniscule}{\@setfontsize\miniscule{5}{6}}
\fi
\makeatother

\DeclareRobustCommand{\optbar}[1]{\shortstack{{\miniscule (\rule[.5ex]{1.25em}{.18mm})}
  \\ [-.7ex] $#1$}}

\def\Kbar    {{\kern 0.2em\overline{\kern -0.2em \PK}{}}\xspace}

\def\KorKbar    {\kern 0.18em\optbar{\kern -0.18em K}{}\xspace}

\def\Dbar    {{\kern 0.2em\overline{\kern -0.2em \PD}{}}\xspace}
\def\D       {{\ensuremath{\PD}}\xspace}

\def\DorDbar    {\kern 0.18em\optbar{\kern -0.18em D}{}\xspace}
\def\Dz      {{\ensuremath{\D^0}}\xspace}

\def\Bbar    {{\ensuremath{\kern 0.18em\overline{\kern -0.18em \PB}{}}}\xspace}

\def\BorBbar    {\kern 0.18em\optbar{\kern -0.18em B}{}\xspace}

  \def\Y#1S{\ensuremath{\PUpsilon{(#1S)}}\xspace}

\def\Lbar        {{\ensuremath{\kern 0.1em\overline{\kern -0.1em\PLambda}}}\xspace}
\def\LorLbar     {\kern 0.18em\optbar{\kern -0.18em \PLambda}{}\xspace}

\def\to                 {\ensuremath{\rightarrow}\xspace}

\def\AT#1     {\ensuremath{A_{\mathrm{T}}^{#1}}\xspace}           

\def\C#1      {\ensuremath{\mathcal{C}_{#1}}\xspace}                       
\def\Cp#1     {\ensuremath{\mathcal{C}_{#1}^{'}}\xspace}                    
\def\Ceff#1   {\ensuremath{\mathcal{C}_{#1}^{\mathrm{(eff)}}}\xspace}        
\def\Cpeff#1  {\ensuremath{\mathcal{C}_{#1}^{'\mathrm{(eff)}}}\xspace}       
\def\Ope#1    {\ensuremath{\mathcal{O}_{#1}}\xspace}                       
\def\Opep#1   {\ensuremath{\mathcal{O}_{#1}^{'}}\xspace}                    

\newcommand{\tev}{\ifthenelse{\boolean{inbibliography}}{\ensuremath{~T\kern -0.05em eV}\xspace}{\ensuremath{\mathrm{\,Te\kern -0.1em V}}}\xspace}
\newcommand{\gev}{\ensuremath{\mathrm{\,Ge\kern -0.1em V}}\xspace}
\newcommand{\mev}{\ensuremath{\mathrm{\,Me\kern -0.1em V}}\xspace}
\newcommand{\kev}{\ensuremath{\mathrm{\,ke\kern -0.1em V}}\xspace}
\newcommand{\ev}{\ensuremath{\mathrm{\,e\kern -0.1em V}}\xspace}
\newcommand{\gevc}{\ensuremath{{\mathrm{\,Ge\kern -0.1em V\!/}c}}\xspace}
\newcommand{\mevc}{\ensuremath{{\mathrm{\,Me\kern -0.1em V\!/}c}}\xspace}
\newcommand{\gevcc}{\ensuremath{{\mathrm{\,Ge\kern -0.1em V\!/}c^2}}\xspace}
\newcommand{\gevgevcccc}{\ensuremath{{\mathrm{\,Ge\kern -0.1em V^2\!/}c^4}}\xspace}
\newcommand{\mevcc}{\ensuremath{{\mathrm{\,Me\kern -0.1em V\!/}c^2}}\xspace}

\def\invfb {\ensuremath{\mbox{\,fb}^{-1}}\xspace}

\def\gsim{{~\raise.15em\hbox{$>$}\kern-.85em
          \lower.35em\hbox{$\sim$}~}\xspace}
\def\lsim{{~\raise.15em\hbox{$<$}\kern-.85em
          \lower.35em\hbox{$\sim$}~}\xspace}

\def\pt         {\mbox{$p_{\rm T}$}\xspace}

\def\tell1  {TELL1\xspace}
\def\ukl1   {UKL1\xspace}

\newcommand{\ie}{\mbox{\itshape i.e.}\xspace}

\def\DzorDzbar                {\kern 0.18em\optbar{\kern -0.18em \Dz}{}\xspace}

\def\Y                        {\ensuremath{\mathcal{N}}\xspace}

\def\pB                       {\ensuremath{p_B}\xspace}

\usepackage{titlesec}
\titlespacing\section{0pt}{10pt}{1pt}
\titlespacing\subsection{0pt}{10pt plus 3pt minus 3pt}{1pt plus 1pt minus 1pt}
\titlespacing\subsubsection{0pt}{8pt plus 3pt minus 3pt}{1pt plus 1pt minus 1pt}

\usepackage{tikz,xcolor,hyperref}

\definecolor{lime}{HTML}{A6CE39}

\title{Partial reconstruction of the decay $B_{s}^{0}\rightarrow \phi(K^{+}K^{-})\eta' (\pi^{+}\pi^{-}\eta)$ at a future high-luminosity $Z$ factory}

\usepackage{background}
\usepackage{xcolor}     

\backgroundsetup{
  scale=1,
  color=gray!90,
  opacity=1,
  angle=0,
  position=current page.south,
  vshift=1.2cm,
  contents={\texttt{*correspondence: jdjuevesano@up.edu.ph}}
}

\usepackage{authblk}

\author{Jeremiah D. Juevesano$^{1,2,3}$}
\author{Marlon P. Brade$^{2,3,4}$}
\author{Dennis C. Arogancia$^{2,3}$}
\author{Jan Mickelle V. Maratas$^{2,3}$}

\affil{$^{1}$Department of Mathematics, Physics, and Computer Science, University of the Philippines - Mindanao, Davao City, Philippines}

\affil{$^{2}$Department of Physics,
Mindanao State University--Iligan Institute of Technology,
Iligan City, Philippines}

\affil{$^{3}$Mindanao Radiation Physics Center, Premier Research Institute of Science and Mathematics, MSU-Iligan Institute of Technology, Iligan City, Philippines}

\affil{$^{4}$Department of Physics,
Western Mindanao State University--Main,
Zamboanga City, Philippines}

\makeatletter
\begin{document}
  
\maketitle

\begin{abstract}
The decay $B_s^0 \to \phi(K^+K^-)\eta'(\pi^+\pi^-\eta)$ is of particular
interest in flavour physics because in the Standard Model it proceeds
predominantly through the quark transition $\bar{b} \to \bar{s}s\bar{s}$.
However, the reconstruction of this decay is experimentally challenging
due to the presence of the neutral $\eta$ particle, which significantly
reduces the overall reconstruction efficiency.

In this study, the feasibility of using the partial reconstruction
technique to reconstruct the invariant mass of the $B_s^0$ meson
without explicitly detecting the $\eta$ particle is investigated.
Monte Carlo simulated events corresponding to an FCC-ee environment
are used, and the effects of detector momentum and vertex resolutions
on the reconstruction performance are examined.

The results show that improved vertex resolution substantially improves
the global reconstructed mass response and partial-reconstruction
efficiency, while improvements in momentum resolution are manifested
primarily through a narrowing of the fitted central DCB component. These findings
suggest that future high-luminosity $e^+e^-$ colliders such as
FCC-ee could effectively exploit the partial reconstruction technique
to study rare decay modes containing undetected neutral particles.
\end{abstract}
\vspace{0.35cm}




\section{Introduction}

In certain cases, a physics analysis is limited by statistics due to 
detector geometry or acceptance, as well as the low reconstruction efficiency of some
particle species, particularly neutral particles such as V$^0$s. In particular, the presence of one or more neutral particles significantly reduces the reconstruction efficiency of analyses targeting such final states. One such decay channel is  $B_{s}^{0}\rightarrow \phi(K^{+}K^{-}) \eta'(\pi^{+}\pi^{-} \eta )$ \footnote{Charge-conjugated processes
are implied in this letter unless stated otherwise.},
where $\eta$ is the neutral final-state particle. Explicitly trying to reconstruct
the $\eta$ particle results in a low overall reconstruction efficiency for this decay mode. Because the 
$\eta$
in the decay chain $ \eta' \rightarrow \pi^{+}\pi^{-} \eta $ is a neutral particle that is difficult to reconstruct in collider detectors, the overall reconstruction efficiency of this decay becomes significantly reduced. This motivates the use of the partial reconstruction technique, which allows the kinematics of the parent $B^0_s$ meson to be determined without explicitly reconstructing the 
$\eta$.
A technique known in the particle physics community as the ``partial
reconstruction technique'' \cite{ms3report, babar_PR}
reconstructs the kinematics of the $B_{s}^0$ meson
even without explicitly reconstructing the $\eta$ particle. The practical application of this technique is currently limited by detector resolution, particularly the momentum and vertex resolutions.  In future particle
colliders however, detector resolution is expected to improve compared to current
technologies. For example, the proposed ILD (International Large Detector) has a momentum resolution of $\Delta p/p$ = 0.1\% + \pt/10$^5$ \gev,
where \pt is the transverse momentum, and primary \& secondary vertex resolutions
of 3 $\mu m$ and 7 $\mu m$, respectively \cite{Behnke:2013lya}. It is expected that
the resolutions of the detector for the
Future Circular Collider (FCC) \cite{Abada:2019lih,Kamenik:2017ghi}, a high
luminosity Z-factory, will be even better than these resolutions by up
to a factor of 4.

\begin{figure}[h]
  \centering
\includegraphics[width=0.7\columnwidth]{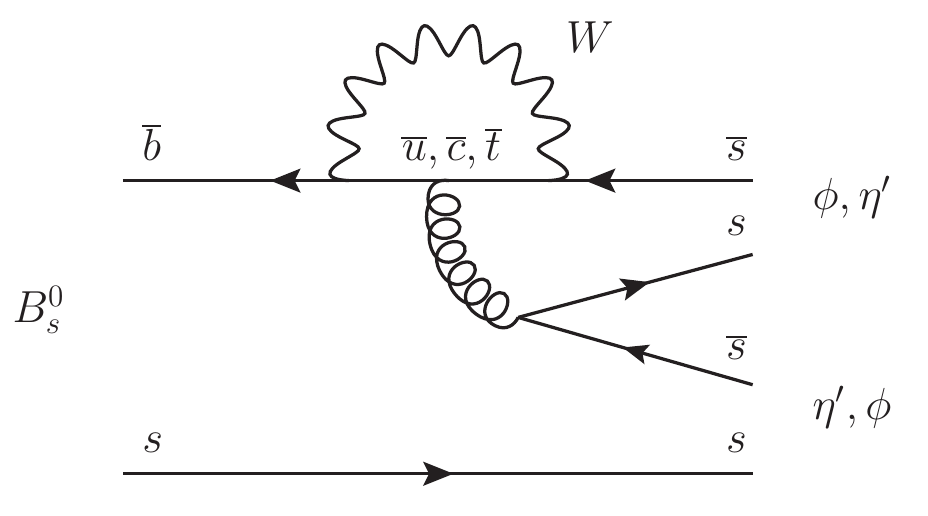}
  \caption{Penguin loop transition of $B_{s}^{0}\rightarrow \phi \eta' $ decay in the SM.}
  \label{loopBsTophietap}
\end{figure}

Beyond the experimental advantages offered by future colliders, this decay channel is also of particular interest from the perspective of flavour physics.
The physics case of the $B_{s}^{0}\rightarrow \phi \eta'$ decay arises from the fact that in the
Standard Model (SM) of particle physics, this decay proceeds predominantly
through the quark transition $\bar{b} \rightarrow \bar{s} s \bar{s}$,
as shown in Figure \ref{loopBsTophietap}. Several studies have predicted the branching fraction
of this decay mode, which ranges from $0.1 \times 10^{-6}$ to $20 \times 10^{-6}$,
with quite large uncertainties due to limited knowledge on the form factors,
penguin contributions, the $\omega-\phi$ mixing angle, and the $s$-quark
mass. Listed in Table \ref{predBsTophietap} are the theoretical predictions for the branching
fraction of the $B_{s}^{0}\rightarrow \phi \eta'$. A better understanding of this value will help
constrain many phenomenological models, particularly in searches for Physics
beyond the SM, \ie New Physics.  A new Flavour Changing Neutral Current (FCNC) could enter inside the loop and hence could enhance the branching fraction of $B_{s}^{0}\rightarrow \phi \eta'$. It has to be noted that this decay mode of $B_{s}^0$ has not yet been
observed to date.

\begin{table}[htbp!]
\centering
\caption{Theoretical predictions for the branching fraction
of $B_s^0 \rightarrow \phi\eta'$ in the SM.}
\label{predBsTophietap}

\renewcommand{\arraystretch}{1.5}

\begin{tabular}{l@{\hspace{1.5cm}}c}
\hline
Theory method & Branching fraction ($\times 10^{-6}$) \\
\hline
QCD factorization \cite{Beneke:2003zv}
    & $0.05^{+1.18}_{-0.19}$ \\
QCD factorization \cite{Cheng:2009mu}
    & $2.2^{+9.4}_{-3.1}$ \\
Perturbative QCD \cite{Ali:2007ff}
    & $0.19^{+0.20}_{-0.13}$ \\
Perturbative QCD \cite{Chen:2007qm}
    & $20.0^{+16.3}_{-9.1}$ \\
SCET \cite{Wang:2008eu}
    & $4.3^{+5.2}_{-3.6}$ \\
SU(3) flavour symmetry \cite{Cheng:2014rfa}
    & $5.5 \pm 1.8$ \\
FAT \cite{Zhou:2016jkv}
    & $13.0 \pm 1.6$ \\
\hline
\end{tabular}
\end{table}

The LHCb collaboration has searched for this decay mode using the proton-proton collision data of LHC, collected at a center-of-mass energies of 7 $TeV$ and 8 $TeV$.
The data corresponds to an integrated luminosity of 3 \invfb. No signal has been
found and hence an upper limit on the branching fraction has been quoted instead,
which is $< 0.82 \times 10^{-6}$ @ 90\% Confidence Level (C.L.) or
$< 1.01 \times 10^{-6}$ @ 95\% C.L \cite{Aaij:2016yuv,Takahashi}.

A proposed future collider, dubbed the Future Circular Collider (FCC) \cite{Abada:2019lih},
will significantly increase the statistics of this decay mode by producing about
10$^{12}$ Z bosons in the first year of data taking in $e^+e^-$ mode (FCC-ee).
FCC, to be hosted by CERN, is a $\sim$80-100 km circular collider to be built
in addition to the existing Large Hadron Collider in order to reach center-of-mass
energies in the $TeV$ scale. Of particular interest in this study is the $e^+e^-$
mode of FCC at a center-of-mass energy of 91 \gev (the mass of the Z boson). In
this mode of operation, 10$^{12}$ Z bosons will decay to $b\bar{b}$ pair about
($15.12 \pm 0.05$)\% of the time, and then the $\bar{b}$-quark (or $b$-quark)
will hadronize to $B_{s}^0$ (or $\bar{B_{s}}$) about ($8.8 \pm 1.3$)\% of the time \cite{Takahashi, pdg_hadronization}.

Using the current experimental upper limit
$\mathcal{B}(B_s^0\to\phi\eta')<8.2\times10^{-7}$ at 90\% C.L.,
together with $\mathcal{B}(Z\to b\bar b)=0.1512$,
$f(b\to B_s^0)=0.088$,
$\mathcal{B}(\phi \to K^+K^-)\simeq 0.49$~\cite{Takahashi}, and
$\mathcal{B}(\eta'\to\eta\pi^+\pi^-)\simeq0.43$ \cite{Takahashi}, the projected
$10^{12}$ $Z$-boson sample corresponds to fewer than approximately
$4.6\times10^3$ signal decays in the final state considered here,
before detector acceptance and reconstruction effects.

Partial reconstruction of
$B_{s}^{0}\rightarrow \phi(K^{+}K^{-})\eta'(\pi^{+}\pi^{-}\eta)$
can increase the accessible signal yield and may improve the prospects
for observing this decay mode in a future high-statistics environment
such as FCC-ee. The increase in statistics
is due to the fact that the $\eta$ particle is not explicitly reconstructed, since the
efficiency of reconstructing this neutral particle is quite low. Instead, the technique
relies on the kinematical and topological constraints to reconstruct the $B^{0}_{s}$. Among the many possible decay modes of $B^{0}_{s}$, this study has chosen the decay
mode to $\phi(K^+ K^-) \eta'(\pi^+\pi^- \eta)$ ($\phi$ decaying to $K^+ K^-$
and $\eta'$ decaying to $\pi^+\pi^- \eta$), since this decay
mode has not yet been observed to date with theoretical prediction on its branching
fraction at the order of $10^{-6}$ \cite{Beneke:2003zv,Cheng:2009mu,Ali:2007ff,Chen:2007qm,Wang:2008eu,Cheng:2014rfa,Zhou:2016jkv}
in the context of the SM. New Physics could however enter in the loop diagram (see
Figure \ref{loopBsTophietap}) that could enhance this branching fraction and hence
this decay mode can serve as a test to the SM.

To simulate the decay events of $B_{s}^{0}\rightarrow \phi(K^{+}K^{-})$ $ \eta'(\pi^{+}\pi^{-} \eta )$ in a FCC-ee environment, \textsc{FCC-Generator} \cite{simkevgen} is used. \textsc{FCC-Generator} is an interface toolkit combining \textsc{PYTHIA} \cite{Sjostrand:2014zea}, a toolkit in simulating high energy primary particles such as $e^+ e^-$ collision, and \textsc{EVTGEN} \cite{Lange:2001uf,Ryd:2005zz}, a software handling the simulation of hadron decays.
The signal decay chain was generated using \textsc{EvtGen}, with the
signal modes forced to $B_s^0\to\phi\eta'$,
$\phi\to K^+K^-$, and $\eta'\to\eta\pi^+\pi^-$.
The corresponding \textsc{EvtGen} decay models were
\texttt{SVS} for the scalar-to-vector--scalar decay
$B_s^0\to\phi\eta'$,
\texttt{VSS} for the vector-to-scalar--scalar decay
$\phi\to K^+K^-$, and
\texttt{PHSP} (phase space) for the three-body decay
$\eta'\to\eta\pi^+\pi^-$.
Thus, the generated decay chain is
\[
B_s^0
\xrightarrow{\texttt{SVS}}
\phi\eta',
\qquad
\phi
\xrightarrow{\texttt{VSS}}
K^+K^-,
\qquad
\eta'
\xrightarrow{\texttt{PHSP}}
\eta\pi^+\pi^-.
\]
These model assignments determine the decay kinematics of the generated
signal events prior to detector smearing and partial reconstruction.
The tracking resolutions achievable at FCC-ee are then incorporated to the simulated events to mimic real detector environment using \textsf{HEPPY-FCC} \cite{heppyfcc}, a lightweight python-based framework handling the overall steering of the event processes. Detector resolutions play a significant role in the determination of the kinematics decay as well as in the measurement of the desired physical quantity. In particular, the detector resolutions such as the vertex and momentum resolutions can significantly affect the partial reconstruction technique used in this study. The momenta of the visible and Monte Carlo (MC) generated final state particles ($K^+ K^-$ from $\phi$ and $\pi^+ \pi^- $ from $\eta'$) are smeared using the equation,
\begin{equation}
  \label{momre}
  \frac{\Delta p}{p} = 0.001 + \frac{\pt}{10^5 \gev}
\end  {equation}
where $\Delta p$ is the resolution, $p$ is the momentum of
the particle, and \pt (in units of \gev) is the particle momentum transverse to the $e^+e^-$
beam direction. These resolutions are approximated from the proposed
detector of the future International Large 
Detector (ILD) \cite{Behnke:2013lya}.

Aside from the momentum smearing of the final-state particles, detector
resolution also affects the reconstructed production and decay vertices of
the $B_s^0$ meson. The baseline vertex-resolution configuration, denoted
by V1, uses production- and decay-vertex resolutions of $3~\mu\mathrm{m}$
and $7~\mu\mathrm{m}$, respectively, motivated by the corresponding ILD
detector resolutions. To study the dependence of the partial reconstruction
on vertex resolution, two additional configurations are considered. In V2
and V4, both the production- and decay-vertex resolutions are improved by
factors of 2 and 4, respectively.

The momentum-resolution configurations are denoted by M1, M2, and M4.
M1 corresponds to the baseline momentum resolution given in Eq.~(1),
while M2 and M4 represent twofold and fourfold improvements, respectively.
For M2 and M4, both terms on the right-hand side of Eq.~(1) are scaled
uniformly by factors of $1/2$ and $1/4$, respectively.

The vertex- and momentum-resolution configurations are combined and
denoted by $\mathrm{V}i\mathrm{M}j$, where $\mathrm{V}i$ specifies the
vertex resolution and $\mathrm{M}j$ specifies the momentum resolution.
The nine configurations considered in this study are summarized in
Table~\ref{tab:resolution_configs}.

\begin{table}[h!]
\centering
\renewcommand{\arraystretch}{1.35}
\caption{Vertex- and momentum-resolution configurations used in this study.
The momentum-resolution scale factor is applied uniformly to both terms
of the baseline resolution given in Eq.~(1).}
\label{tab:resolution_configs}

\begin{tabular}{cccc}
\hline
Configuration &
Production vertex &
Decay vertex &
Momentum-resolution scale \\
\hline
V1M1 & $3~\mu\mathrm{m}$   & $7~\mu\mathrm{m}$   & $1$   \\
V1M2 & $3~\mu\mathrm{m}$   & $7~\mu\mathrm{m}$   & $1/2$ \\
V1M4 & $3~\mu\mathrm{m}$   & $7~\mu\mathrm{m}$   & $1/4$ \\
\hline
V2M1 & $3/2~\mu\mathrm{m}$ & $7/2~\mu\mathrm{m}$ & $1$   \\
V2M2 & $3/2~\mu\mathrm{m}$ & $7/2~\mu\mathrm{m}$ & $1/2$ \\
V2M4 & $3/2~\mu\mathrm{m}$ & $7/2~\mu\mathrm{m}$ & $1/4$ \\
\hline
V4M1 & $3/4~\mu\mathrm{m}$ & $7/4~\mu\mathrm{m}$ & $1$   \\
V4M2 & $3/4~\mu\mathrm{m}$ & $7/4~\mu\mathrm{m}$ & $1/2$ \\
V4M4 & $3/4~\mu\mathrm{m}$ & $7/4~\mu\mathrm{m}$ & $1/4$ \\
\hline
\end{tabular}
\end{table}

In order to study the intrinsic performance of the partial reconstruction
technique and its dependence on detector resolution, approximately
110,000 $B_{s}^{0}\rightarrow \phi(K^{+}K^{-})$ $ \eta'(\pi^{+}\pi^{-} \eta )$ signal MC events were generated and partially
reconstructed. This enlarged simulated sample is not intended to represent
the expected FCC-ee signal yield; rather, it provides sufficient Monte
Carlo statistics to characterize the reconstructed mass response and to
compare the different reconstruction and detector-resolution scenarios
with reduced statistical fluctuations.

\section{Partial Reconstruction Methodology}

\begin{figure}[h!]
  \centering
  \includegraphics[width=0.8\columnwidth]{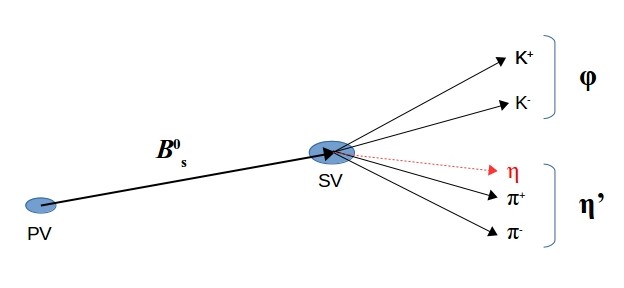}
  \caption{Typical topology of the $B_{s}^{0}\rightarrow \phi(K^{+}K^{-})$ $ \eta'(\pi^{+}\pi^{-} \eta )$,
           where the vertices are reconstructed and the charged tracks
           are detectable.}
  \label{partialrecon}
\end{figure}

The topology of the
$B_{s}^{0}\rightarrow \phi(K^{+}K^{-})$ $ \eta'(\pi^{+}\pi^{-} \eta )$ decay is characterized by two vertices
(primary and secondary) and two oppositely-charged kaons
coming from the secondary vertex whose combined mass points to
the nominal mass of $\phi$, and two oppositely-charged pions
coming from the same secondary vertex as shown in Figure \ref{partialrecon}. The four-vector
energy-momentum of the $B^{0}_{s}$ can be determined without explicitly
detecting the neutral $\eta$ particle as the decay is sufficiently
constrained by the topological and kinematical constraints.

There are a total of seven (7) degrees of freedom (\ie unknowns) \textit{a priori},
which are the energy and 3-vector momentum of $B^{0}_{s}$ and the 3-vector
momentum of the undetected $\eta$ particle. Hence, a total of at
least seven (7) constraints are necessary to solve the decay.
The first four constraints come from the four-vector momentum
conservation of the decay, \ie,

\begin{equation}
 \overbrace{\sqrt{p_{B}^{2} + m_{B}^{2}}}^{\ensuremath{E_B}\xspace}
 =
 \overbrace{\sqrt{p_{\phi}^{2} + m_{\phi}^{2}}}^{\ensuremath{E_{\phi}}\xspace}
 +
 \overbrace{\sqrt{p_{2\pi}^{2} + m_{2 \pi}^{2}}}^{\ensuremath{E_{2\pi}}\xspace}
 +
 \overbrace{\sqrt{p_{\eta}^{2} + m_{\eta}^{2}}}^{\ensuremath{E_{\eta}}\xspace}
\end{equation}

\begin{equation}
  \overrightarrow{p_B}
  =
  \overrightarrow{p_{\phi}}
  +
  \overrightarrow{p_{2\pi}}
  +
  \overrightarrow{p_{\eta}}~,
\end{equation}

where $\overrightarrow{p_B}$, $\overrightarrow{p_{\phi}}$,
$\overrightarrow{p_{2\pi}}$ and $\overrightarrow{p_{\eta}}$ are the corresponding three-vector momenta,
while $m$ denotes the corresponding invariant or fixed particle
mass as specified below.

Two additional constraints come from the fact that the four charged
tracks ($K^+ K^- \pi^+ \pi^-$) originate from a common secondary
vertex allowing the determination of the direction of
$\overrightarrow{\pB}$, hence,

\begin{equation}
  \overrightarrow{p_B}
  =
  p_{\phi}^{\parallel}
  +
  p_{2\pi}^{\parallel}
  +
  p_{\eta}^{\parallel}
\end{equation}

\begin{equation}
  \overrightarrow{0}
  =
  p_{\phi}^{\perp}
  +
  p_{2\pi}^{\perp}
  +
  p_{\eta}^{\perp} ~ ,
\end{equation}

where $ p_{\phi}^{\parallel}$, $p_{2\pi}^{\parallel}$
and $p_{\eta}^{\parallel}$ are vector components parallel
(or anti-parallel) to the direction of $\overrightarrow{p_B}$;
while $p_{\phi}^{\perp}$, $p_{2\pi}^{\perp}$ and
$p_{\eta}^{\perp}$ are the vector components of the said
particles perpendicular to $\overrightarrow{p_B}$.

The final constraint is obtained by requiring that
the combined mass of the two pions and the undetected particle
$\eta$ be equal to the nominal mass of $\eta'$, \ie $m_{\eta'}$.
Again using the conservation of four-vector momentum of
$B_{s}^{0} \rightarrow \phi \eta' (\pi^{+}\pi^{-} \eta )$, hence,

\begin{equation}
  m_{B}^2
  =
  m_{\phi}^2
  +
  m_{\eta'}^2
  +
  2(E_{\phi} E_{\eta'}
  -
  \overrightarrow{p_{\phi}}\cdot\overrightarrow{p_{\eta'}})~,
\end{equation}

where
$E_{\phi} = \sqrt{p_{\phi}^2 + m_{\phi}^2}$ and

\[
E_{\eta'}
=
E_{2\pi} + E_{\eta}
=
\sqrt{p_{2\pi}^2+ m_{2\pi}^{2}}
+
\sqrt{p_{\eta}^2 + m_{\eta}^2}
\]

are the energies of $\phi$ and $\eta'$, respectively while

\[
\overrightarrow{p_{\eta'}}
=
\overrightarrow{p_{2\pi}}
+
\overrightarrow{p_{\eta}}
\]

is the three-vector momentum of $\eta'$.

The fixed particle masses used in the reconstruction are
$m_{K^\pm}=493.677$ MeV,
$m_{\pi^\pm}=139.57039~\mathrm{MeV}$,
$m_{\eta}=547.862~\mathrm{MeV}$, and
$m_{\eta'}=957.78~\mathrm{MeV}$.
The $\phi$-candidate and dipion invariant masses,
$m_{\phi}$ and $m_{2\pi}$, are reconstructed event by event
from the corresponding visible four-momenta. In particular,
no fixed nominal $\phi$ mass enters the quadratic solution in
Eqs.~(7)--(9). The $B_s^0$ mass is reconstructed and is not
imposed as a constraint in the experimentally applicable
reconstruction. The system of equations generally leads to quadratic solutions for
the missing momentum of the $\eta$ particle. The four constraints
from the conservation of four-vector momentum, two constraints
arising from the determination of the $B^{0}_{s}$ flight direction
from the reconstructed primary and secondary vertices, and the
invariant mass constraint of the $\eta'$ together provide seven
independent equations. These match the seven unknown parameters
of the system. The seven constraints therefore allow the seven
unknowns to be solved, albeit leading to quadratic equations. 

The resulting quadratic ambiguity constitutes the principal challenge
of the partial reconstruction algorithm, since only one of the two
quadratic solutions corresponds to the physically correct event
kinematics. Consequently, an additional reconstruction criterion is
required to determine which solution is retained for each event.

\subsection{Reconstruction Procedure and Solution Selection}

The quadratic solutions arise from solving for the missing momentum
$p_{\eta}$, resulting in

\begin{equation}
\vec{p}_{\eta}^{,\parallel}
=
\left( A \pm \sqrt{B} \right)\hat{p}_{B},
\end{equation}

where

\begin{equation}
A =
\frac{
\left[
(m^{2}_{\eta'} - m^{2}_{2\pi} -m^{2}_{\eta}  )
-
2\vec{p}_{2\pi}^{,\perp}\cdot
\left(
\vec{p}_{2\pi}^{,\perp}
+
\vec{p}_{\phi}^{,\perp}
\right)
\right]
p_{2\pi}^{,\parallel}
}{
2\left(
p_{2\pi}^{,\perp 2}
+
m_{2\pi}^{2}
\right)
},
\end{equation}

and

\begin{equation}
B =
\left\{
\frac{
\left[
(m^{2}_{\eta'} - m^{2}_{2\pi} -m^{2}_{\eta}  )
-
2\vec{p}_{2\pi}^{,\perp}\cdot
\left(
\vec{p}_{2\pi}^{,\perp}
+
\vec{p}_{\phi}^{,\perp}
\right)
\right]^{2}
}{
4\left(
p_{2\pi}^{,\perp 2}
+
m_{2\pi}^{2}
\right)^{2}
}
-
\frac{
m_{\eta}^{2}
+
\left(
\vec{p}_{\phi}^{,\perp}
+
\vec{p}_{2\pi}^{,\perp}
\right)^{2}
}{
p_{2\pi}^{,\perp 2}
+
m_{2\pi}^{2}
}
\right\}
E_{2\pi}^{2}.
\end{equation}


For each event, the two possible quadratic solutions are evaluated.
For each solution, the corresponding reconstructed four-momentum of the
\(B_s^0\) candidate is obtained. The physical quadratic solution is then selected as the
solution whose reconstructed \(B_s^0\) energy is closest to the nominal beam
energy,

\[
E_{\rm beam}=\frac{\sqrt{s}}{2}
=\frac{91.188~{\rm GeV}}{2}
=45.594~{\rm GeV}.
\]

This criterion is used as an approximate kinematic guide for resolving the
quadratic ambiguity in an $e^+e^-$ collider operating at the $Z$ pole.
The selection relies only on reconstructed event kinematics and therefore
provides a practical procedure for resolving the quadratic ambiguity
without requiring knowledge of the true missing-particle momentum.
It does not require each
$B_s^0$ meson to carry exactly the beam energy, since its actual energy
depends on the fragmentation of the parent $b$ quark. Solutions producing unphysical kinematics, such as negative energies,
imaginary momentum components, NaN values, or other non-reconstructible
configurations, are rejected.
The accepted reconstructed candidates are required to satisfy

\[
4700 < m_{B_s^0} < 6000~{\rm MeV},
\]

which defines a sufficiently broad mass window to retain the reconstructed
signal while suppressing highly unphysical kinematic solutions. The
reconstruction criterion described above defines the experimentally
accessible partial reconstruction procedure adopted throughout this work.
Before detector effects are introduced, its performance is validated
independently using MC-truth events, as presented in the
following subsection.

\subsection{MC-Truth Validation of the Partial Reconstruction Algorithm}

The reconstruction criterion introduced in the previous subsection provides a practical procedure for resolving the quadratic ambiguity using only reconstructed event information. Before detector effects are introduced, its intrinsic performance is validated using MC-truth events, allowing the reconstruction algorithm to be assessed independently of detector resolution.

To assess the reconstruction procedure, three treatments of the quadratic
ambiguity are considered. The first corresponds to the \emph{randomized}
reconstruction, in which one of the two quadratic solutions is selected
with equal probability. This provides an unbiased reference describing the
expected performance in the absence of any kinematic selection criterion.

The second corresponds to the \emph{selected} reconstruction described in
Section~2.1, where the quadratic solution whose reconstructed $B_s^0$
energy is closest to the nominal beam energy is retained. This represents
the experimentally applicable reconstruction strategy adopted throughout
this work.

Finally, an idealized \emph{cheated} reconstruction is introduced for
validation purposes. In this case, the quadratic solution whose
reconstructed $B_s^0$ energy is closest to

\[
E_{B_s^0}
=
\sqrt{m_{B_s^0}^{2}+p_{B_s^0}^{2}}
\]

is selected, where the $B_s^0$ momentum is reconstructed using the true
MC-truth momentum of the missing $\eta$ particle. Since this information is unavailable during reconstruction, the
cheated solution does not represent an experimentally realizable
reconstruction strategy. Instead, it serves as an ideal benchmark for
the partial reconstruction algorithm. By selecting the physically
correct solution of the quadratic ambiguity, it represents the best
achievable reconstruction performance and therefore quantifies the
degradation introduced by the unresolved ambiguity. The randomized and
selected reconstruction strategies are subsequently evaluated according
to how closely they approach this benchmark.

The validation is performed using simulated
$B_s^0\rightarrow\phi(K^+K^-)\eta'(\eta\pi^+\pi^-)$
signal events produced in
$e^+e^-\rightarrow Z\rightarrow b\bar{b}$ collisions at the
$Z$ pole. Although the projected FCC-ee dataset is expected to yield
only $\mathcal{O}(10^3)$ signal events before detector effects, the
validation studies presented in this work employ an independent
high-statistics MC-truth sample. The purpose and size of this sample
are discussed below.

To obtain statistically stable distributions for validating the partial
reconstruction algorithm and determining the parameters of the empirical
model, a larger MC-truth sample containing approximately
$1.1\times10^5$ signal events is used throughout this section. This
enlarged sample is employed solely to reduce statistical fluctuations in
the validation of the reconstruction algorithm and does not represent
the expected event yield for the projected FCC-ee dataset. Using the larger simulated sample reduces statistical fluctuations and
provides a more stable characterization of the reconstruction algorithm
and its reconstructed mass response.

The reconstructed invariant-mass distributions obtained from these
high-statistics samples are subsequently characterized using an
empirical parameterization to enable a quantitative comparison of the
different reconstruction strategies.

\subsubsection{Empirical Characterization of the Reconstructed Mass Response}

The reconstructed invariant-mass distributions obtained from the
randomized, selected, and cheated reconstruction strategies exhibit
asymmetric and non-Gaussian features arising from the nonlinear
kinematic constraints imposed by the partial reconstruction algorithm.
Consequently, an empirical parameterization capable of describing both
the narrow reconstructed core and the asymmetric tails is required.
Throughout this work, these reconstructed mass distributions are
parameterized using a Double Crystal Ball (DCB) function \cite{cbcb}.

The DCB combines a Gaussian core, characterized by the peak position
$\mu$ and width $\sigma$, with independent left- and right-sided
power-law tails parameterized by
$(\alpha_L,n_L)$ and $(\alpha_R,n_R)$. The Gaussian core parameterizes the central reconstructed mass component,
while the two independent power-law tails provide additional flexibility
to describe the broader asymmetric structure. In detector-level analyses, the low-mass tail commonly accommodates energy-loss effects such as imperfect momentum reconstruction or bremsstrahlung, whereas the high-mass tail empirically describes events reconstructed above the central peak. In the present study, the DCB tails are used purely as an empirical parameterization of the reconstructed mass response and are not interpreted as arising from any single physical process.
This flexibility enables the model to describe the asymmetric
reconstructed mass response produced by the partial reconstruction
algorithm.

The DCB is employed solely as an empirical parameterization of the reconstructed mass response rather than as a model of the intrinsic $B_s^0$ resonance line shape.
Consequently, the fitted DCB parameters are interpreted as descriptors
of the reconstruction algorithm rather than physical properties of the
$B_s^0$ resonance. The same empirical parameterization is retained for
both the MC-truth and detector-smeared samples to ensure that any
differences observed between the two arise from detector effects rather
than from a change in the statistical model used to describe the
reconstructed mass distributions.

The DCB parameters are determined from an unbinned maximum likelihood fit implemented in the \textsc{RooFit} framework. Histogram binning is introduced only after the fit for visualization and for evaluating the goodness of fit, and therefore does not affect the estimation of the DCB parameters.
The agreement between the fitted DCB model and the simulated mass distribution is quantified using the Poisson deviance~\cite{Bonamente2024},
\begin{equation}
D =
2\sum_{i=1}^{N_{\rm bin}}
\left[
E_i-N_i+
N_i\ln\left(\frac{N_i}{E_i}\right)
\right],
\end{equation}
where $N_i$ is the observed number of events in bin $i$ and $E_i$ is the corresponding expected number of events from the fitted DCB model. For an empty observed bin, $N_i=0$, the logarithmic term is defined by its limiting value, $N_i\ln(N_i/E_i)=0$, so that the bin contributes $2E_i$ to the deviance.

The expected count in each bin is obtained by integrating the fitted DCB probability density over the bin,
\begin{equation}
E_i =
N_{\rm evt}
\frac{
\displaystyle\int_{m_i^{\rm low}}^{m_i^{\rm high}}
f(m;\hat{\boldsymbol{\theta}}),dm
}{
\displaystyle\int_{m_{\rm min}}^{m_{\rm max}}
f(m;\hat{\boldsymbol{\theta}}),dm
},
\end{equation}
where $N_{\rm evt}$ is the total number of reconstructed events in the fitted mass range, $f(m;\hat{\boldsymbol{\theta}})$ is the DCB probability density evaluated using the parameters obtained from the unbinned fit, and $[m_i^{\rm low},m_i^{\rm high}]$ denotes the boundaries of bin $i$. This construction ensures that the total expected event count is normalized to the observed sample size.

The Poisson deviance is used because the histogram entries are integer event counts and therefore naturally follow a count-data description. Unlike a Gaussian $\chi^2$ statistic, it does not require assigning Gaussian uncertainties to the individual bins and remains well defined for empty observed bins. This is particularly useful when the occupancy varies substantially across the reconstructed mass distribution.

For comparison among fits, the nominal number of degrees of freedom is defined as
\begin{equation}
ndf=N_{\rm bin}-N_{\rm par}-1,
\end{equation}
where $N_{\rm par}=6$ is the number of floating DCB shape parameters and the additional subtraction accounts for the fixed total-event normalization. The ratio $D/ndf$ is reported as a post-fit goodness-of-fit diagnostic.

\begin{figure}[htbp]
\centering
\includegraphics[width=0.99\textwidth]{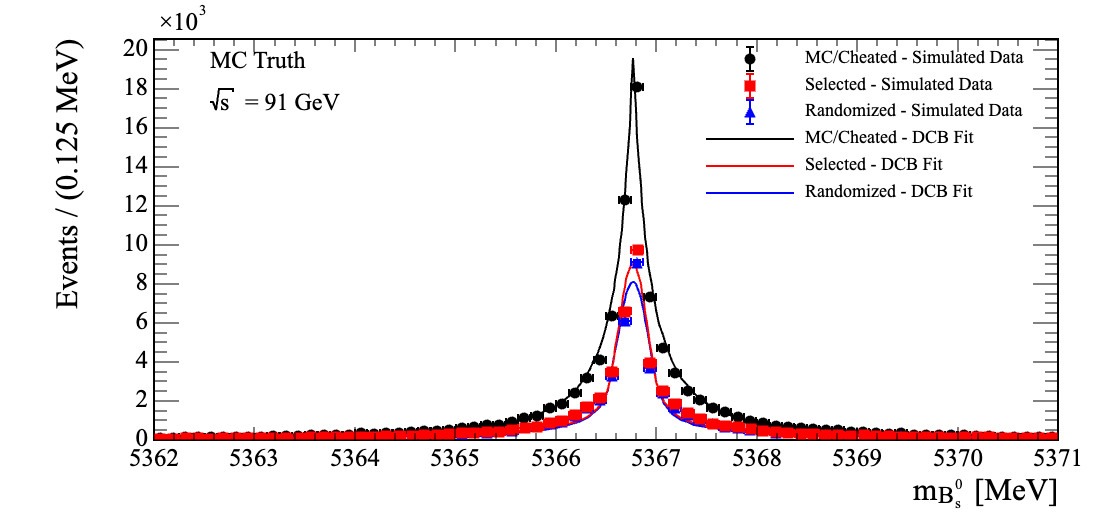}
\caption{
Reconstructed $B_s^0$ invariant-mass distributions obtained from the
cheated (black), selected (red), and randomized (blue)
reconstruction strategies using MC-truth events.
The overlaid curves correspond to the unbinned maximum-likelihood
Double Crystal Ball (DCB) fits.
The fits are performed over the interval
$4700 < m_{B_s^0} < 6000~\mathrm{MeV}$,
while only the vicinity of the reconstructed peak is displayed for
visual clarity.
}
\label{fig:validation_dcb}
\end{figure}

Figure~\ref{fig:validation_dcb} illustrates the reconstructed
$B_s^0$ invariant-mass distributions together with their corresponding
DCB parameterizations. The cheated reconstruction exhibits the
narrowest reconstructed peak, whereas the randomized and selected
strategies display broader asymmetric mass responses arising from the
remaining quadratic ambiguity. Visually, the selected reconstruction
produces a more localized central peak than the randomized approach.
The quantitative comparison of these distributions is presented in the
following subsection using both the fitted DCB parameters and
model-independent statistical measures.

\subsubsection{Selection of the Reconstruction Strategy}

The cheated, randomized, and selected reconstruction strategies are
characterized using both the parameters of the fitted double-sided Crystal
Ball (DCB) function and quantities evaluated directly from the reconstructed
$B_s^0$ invariant-mass distributions. The corresponding DCB parameters
are summarized in Table~\ref{tab:dcb_fit_parameters}, while the RMS,
central 68\% width, and partial-reconstruction efficiency are reported in
Table~\ref{tab:rms_w68_eff}. The post-fit goodness-of-fit diagnostics
based on the Poisson deviance are presented separately in
Table~\ref{tab:poisson_deviance}. The choice of reconstruction strategy is based on the
model-independent reconstruction measures together with the
experimentally applicable solution-selection prescription.

A total of $110{,}000$ signal events were initially generated. During
validation of the generated sample, a subset of events was found to be
affected by an internal event-generation issue and was intentionally
removed prior to the reconstruction analysis. This left $109{,}016$
valid generated events, which form the common input sample used
throughout the revised reconstruction study. All partial-reconstruction
efficiencies reported below are therefore evaluated relative to these
$109{,}016$ events.

The fitted DCB peak position, $\mu$, is consistent among the three
reconstruction strategies, with
$\mu\simeq5366.77~\mathrm{MeV}$. Thus, the treatment of the quadratic
ambiguity has little effect on the position of the central reconstructed
peak. The fitted DCB core-width parameters are
$\sigma=0.00132\pm0.00045~\mathrm{MeV}$ for the cheated reconstruction,
$\sigma=0.16417\pm0.00344~\mathrm{MeV}$ for the randomized
reconstruction, and
$\sigma=0.15314\pm0.00325~\mathrm{MeV}$ for the selected
reconstruction. The parameter $\sigma$ describes only the narrow central
component of the empirical DCB function and is therefore not interpreted
as the width of the complete reconstructed mass distribution. This
distinction is particularly important for the cheated reconstruction,
for which the fitted DCB core is much narrower than the overall mass
response. The fitted $\sigma$ values therefore provide a characterization of the
central DCB component, while the overall reconstruction performance is
assessed using the RMS, $W_{68}$, and partial-reconstruction efficiency.

To characterize the reconstructed distributions independently of the
DCB parameterization, the root-mean-square (RMS) and the width of the
central 68\% interval, $W_{68}$, are evaluated directly from the
reconstructed masses within the common
$4700$--$6000~\mathrm{MeV}$ analysis window. For a sample containing
$N$ accepted candidates with reconstructed invariant masses $m_i$, the
RMS is defined as

\begin{equation}
    \mathrm{RMS}
    =
    \sqrt{
    \frac{1}{N}
    \sum_{i=1}^{N}
    \left(m_i-\bar{m}\right)^2
    },
    \qquad
    \bar{m}
    =
    \frac{1}{N}
    \sum_{i=1}^{N}m_i .
\end{equation}

The central 68\% width is defined through the 16th and 84th percentiles
of the reconstructed mass distribution,

\begin{equation}
    W_{68}=q_{84}-q_{16},
\end{equation}

where $q_{16}$ and $q_{84}$ denote the corresponding mass quantiles.
The RMS is sensitive to the overall spread and to events in the tails,
whereas $W_{68}$ provides a more robust characterization of the central
part of the reconstructed mass distribution.

For the cheated reconstruction, the RMS and $W_{68}$ are
$5.125~\mathrm{MeV}$ and $1.781~\mathrm{MeV}$, respectively. The
randomized strategy gives an RMS of $168.638$ MeV and
$W_{68}=128.906~\mathrm{MeV}$, while the selected strategy gives
$160.923~\mathrm{MeV}$ and $106.614~\mathrm{MeV}$, respectively.
The substantially larger widths of the randomized and selected samples
relative to the cheated benchmark reflect the broader reconstructed
distributions obtained when the quadratic ambiguity must be resolved
without generator-level information. Among the two experimentally
applicable strategies, the selected reconstruction gives both a smaller
RMS and a smaller $W_{68}$ than random solution selection, indicating
a more localized reconstructed mass distribution.

The partial-reconstruction efficiency is defined as

\begin{equation}
    \epsilon_{\mathrm{PR}}
    =
    \frac{N_{\mathrm{acc}}}
         {N_{\mathrm{gen}}}
    \times 100\%,
\end{equation}

where $N_{\mathrm{gen}}=109{,}016$ is the number of valid generated
events and $N_{\mathrm{acc}}$ is the number of events for which the
partial reconstruction succeeds and the reconstructed invariant mass
satisfies

\begin{equation}
    4700 < m_{B_s^0} < 6000~\mathrm{MeV}.
\end{equation}

The cheated reconstruction retains $105,818$ of the $109{,}016$ valid
generated events, corresponding to a partial-reconstruction efficiency of
$97.07\%$. Among the experimentally realizable approaches, the randomized
strategy retains $98{,}792$ events ($90.62\%$), while the selected strategy
retains $100{,}590$ events ($92.27\%$). Thus, the selected prescription
retains $1{,}798$ more events than the randomized treatment while also
giving smaller RMS and $W_{68}$ values.

The origin of the non-unit efficiency and finite width of the cheated
MC-truth reconstruction was investigated through dedicated closure tests.
Of the $109{,}016$ generated events, $105{,}818$ produced real quadratic
solutions when the fixed nominal value
$m_{\eta'}=957.78~\mathrm{MeV}$ was imposed, while $3{,}198$ events
failed at the quadratic-solution stage. When the event-by-event generated
$\pi^+\pi^-\eta$ invariant mass was used instead in a truth-only closure
test, the number of events without a real quadratic solution decreased
from $3{,}198$ to $62$. Over the same test, $W_{68}$ decreased from
$1.784~\mathrm{MeV}$ to $0.00137~\mathrm{MeV}$. This identifies the
difference between the fixed nominal $\eta'$ mass and its generated
event-by-event value as the dominant source of the MC-truth reconstruction
loss and finite central width. The generated $\eta'$ mass is used only
for this closure test; the physical reconstruction continues to use the
nominal $\eta'$ mass.

The DCB parameters are determined from unbinned maximum-likelihood
fits. Histogram binning is introduced only for visualization and for
the post-fit goodness-of-fit evaluation and therefore does not affect
the fitted DCB parameters. Figure~3 is displayed using a bin width of
$0.125~\mathrm{MeV}$, corresponding to the finest binning considered
in the goodness-of-fit study of the MC reconstruction samples.

The post-fit goodness-of-fit is evaluated using the Poisson deviance,
which compares the observed and expected event counts without requiring
a Gaussian approximation for the bin-count uncertainties. For the MC
reconstruction samples, the diagnostic is evaluated using bin widths of
$0.125$, $0.250$, and $0.500~\mathrm{MeV}$ over the
$4700$--$6000~\mathrm{MeV}$ interval. These correspond to
$10{,}400$, $5{,}200$, and $2{,}600$ histogram bins and hence to
$10{,}393$, $5{,}193$, and $2{,}593$ degrees of freedom, respectively,
using $ndf=N_{\rm bin}-6-1$.

The three histogram scales are retained to show explicitly how the
post-fit diagnostic changes with binning. The finer binnings are
sensitive to local structure but contain a larger number of
low-occupancy bins, whereas the coarser binning probes agreement over
a broader mass scale. The resulting $D/\mathrm{ndf}$ values are
therefore interpreted as binning-dependent post-fit diagnostics rather
than as a single measure of fit adequacy. The randomized reconstruction
gives $D/\mathrm{ndf}=1.654$, $2.064$, and $2.954$, while the selected
reconstruction gives $1.679$, $2.163$, and $3.146$ for the three bin
widths, respectively. For the cheated sample, the corresponding values
are $0.468$, $0.715$, and $1.210$.

The variation of $D/\mathrm{ndf}$ with histogram bin width further
shows that these quantities should not be used as absolute measures of
fit quality or as criteria for choosing among the reconstruction
strategies. The choice between the randomized and selected prescriptions is therefore
based on the RMS, $W_{68}$, partial-reconstruction efficiency, and the
deterministic nature of the experimentally applicable solution-selection
criterion.

All three unbinned DCB fits converge successfully, with fit status equal
to zero, covariance quality equal to three, and small
estimated-distance-to-minimum (EDM) values. These quantities indicate
stable numerical minimization and well-determined fitted covariance
matrices, but do not by themselves establish that the DCB fully describes
the reconstructed mass distributions. The Poisson-deviance results show
residual differences between the empirical DCB parameterization and the
full reconstructed spectra, particularly for the randomized and selected
samples, for which $D/\mathrm{ndf}$ increases with bin width. The DCB is
therefore used as an empirical characterization of the reconstructed mass
response rather than as an exact statistical description of the full
distribution.

Among the two experimentally applicable treatments of the quadratic
ambiguity, the selected strategy gives a smaller RMS and central 68\%
width together with a larger partial-reconstruction efficiency than random
solution selection. It also provides a deterministic prescription by
selecting the quadratic solution whose reconstructed kinematics are
closest to the expected beam-energy configuration. On the basis of these
model-independent reconstruction measures and its deterministic
experimental prescription, the selected reconstruction strategy is adopted
for the subsequent detector-resolution study.

\section{Results and Discussion}

The MC-truth validation presented in the previous section established
the selected reconstruction strategy as the experimentally realizable
treatment adopted for resolving the quadratic ambiguity. The analysis
is next extended to detector-level samples by incorporating the
ILD-like tracking and vertex resolutions implemented within the
HEPPY-FCC simulation framework. The purpose of this study is to examine
how the partial-reconstruction performance changes as the assumed
vertex and charged-particle momentum resolutions are varied.

The baseline detector configuration is denoted V1M1. The V1
configuration corresponds to the baseline primary- and
secondary-vertex resolutions of $3~\mu\mathrm{m}$ and
$7~\mu\mathrm{m}$, respectively, as listed in Table~2, while V2 and V4
denote two- and four-fold improvements of the vertex resolution. The M1
configuration corresponds to the baseline charged-particle momentum
resolution of Eq.~(1), while M2 and M4 denote two- and four-fold
improvements of the momentum resolution relative to M1.

All reconstructed mass distributions are analyzed within the common
$4700$--$6000~\mathrm{MeV}$ interval. The DCB parameters are obtained
from unbinned maximum-likelihood fits, whereas the RMS and $W_{68}$ are
evaluated directly from the reconstructed candidates within the same
mass window. Histogram binning therefore does not affect the fitted DCB
parameters. For visualization, the detector-level distributions are
shown using $5~\mathrm{MeV}$ bins. In all comparison figures, each
distribution is normalized independently to the number of candidates
in the corresponding sample. The post-fit goodness-of-fit is evaluated separately using the Poisson
deviance for bin widths of $0.5$, $5$, and $10~\mathrm{MeV}$ over the
same $4700$--$6000~\mathrm{MeV}$ interval. These correspond to
$2600$, $260$, and $130$ histogram bins, respectively. With six fitted
DCB shape parameters and one normalization constraint, the corresponding
numbers of degrees of freedom are
$ndf=N_{\rm bin}-6-1=2593$, $253$, and $123$.
The three bin widths are used to examine the stability of the post-fit
description over different histogram scales: the $0.5~\mathrm{MeV}$
binning probes finer structure in the reconstructed distribution,
whereas the $5$ and $10~\mathrm{MeV}$ binnings reduce the influence of
low-occupancy bins and test the description of the broader spectral
shape. The dependence of $D/\mathrm{ndf}$ on these choices is therefore
reported explicitly rather than interpreting a single binning as a
definitive measure of fit quality.

Before studying the variation of the detector resolutions, the cheated,
randomized, and selected reconstruction strategies are reevaluated
under the common V1M1 detector configuration. This provides a
detector-level counterpart to the MC-truth comparison and allows the
behavior of the three reconstruction prescriptions to be examined after
detector smearing.

\begin{figure}[htbp]
\centering
\includegraphics[width=0.99\textwidth]{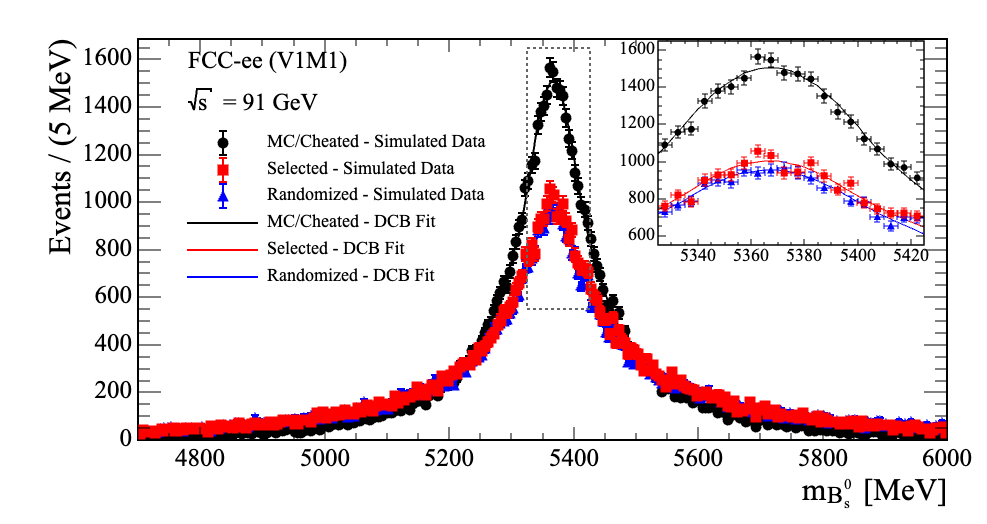}
\caption{
Detector-smeared reconstructed $B_s^0$ invariant-mass distributions
for the cheated, selected, and randomized reconstruction strategies
using the baseline V1M1 detector configuration. The curves show the
corresponding DCB functions obtained from unbinned maximum-likelihood
fits.
}
\label{fig:ILDBaseline}
\end{figure}

Figure~4 shows the reconstructed $B_s^0$ invariant-mass distributions
for the three reconstruction strategies. The fitted DCB core-width
parameters are $47.779\pm1.707~\mathrm{MeV}$ for the cheated
reconstruction, $50.369\pm3.197~\mathrm{MeV}$ for the selected
reconstruction, and $52.371\pm3.080~\mathrm{MeV}$ for the randomized
reconstruction. The overall reconstruction performance is further
characterized using the RMS, $W_{68}$, and partial-reconstruction
efficiency evaluated directly from the reconstructed event samples.

The model-independent measures provide a more direct characterization
of the overall reconstruction performance. The
cheated reconstruction gives an RMS of $164.314~\mathrm{MeV}$ and
$W_{68}=234.139~\mathrm{MeV}$. The selected reconstruction gives
$223.201~\mathrm{MeV}$ and $376.018~\mathrm{MeV}$, respectively,
whereas the randomized reconstruction gives $228.853~\mathrm{MeV}$
and $391.826~\mathrm{MeV}$. Relative to the $109{,}016$ valid generated events, the cheated,
selected, and randomized V1M1 reconstructions retain $62{,}693$,
$58{,}950$, and $57{,}525$ events, corresponding to
partial-reconstruction efficiencies of $57.51\%$, $54.07\%$, and
$52.77\%$, respectively. The selected reconstruction therefore retains
$1{,}425$ more events than random selection and gives smaller RMS and
$W_{68}$ values, consistent with the behavior observed in the MC-truth
study.

The selected reconstruction is consequently retained for the
detector-resolution study below. The post-fit Poisson-deviance values
for the three V1M1 reconstruction strategies are used only as
diagnostics of the empirical DCB description and not as criteria for
selecting the reconstruction strategy.

\begin{figure}[htbp]
\centering
\includegraphics[width=0.99\textwidth]{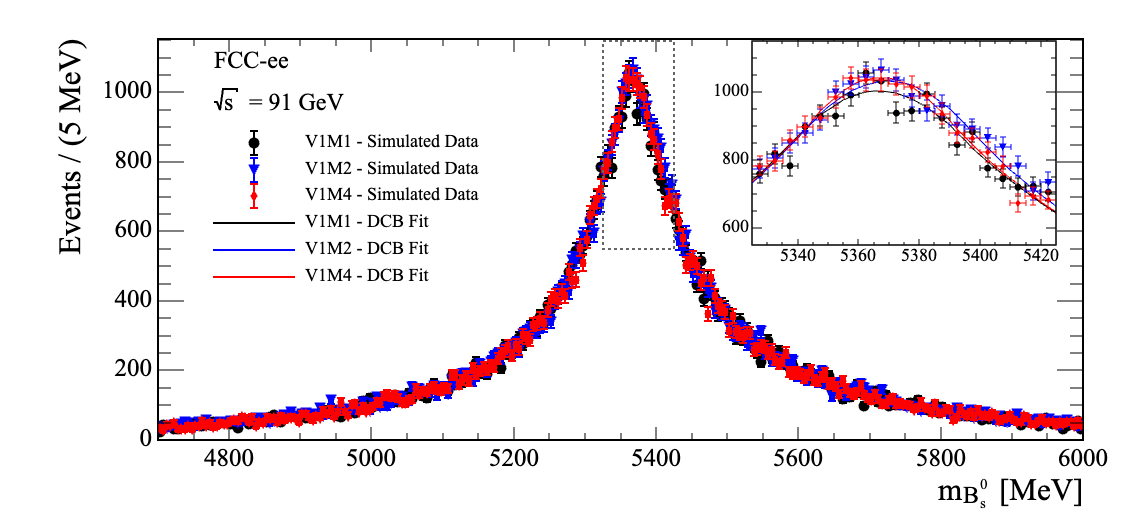}
\caption{
Reconstructed $B_s^0$ invariant-mass distributions and corresponding
DCB fits for the V1M1, V1M2, and V1M4 configurations using the
selected reconstruction strategy.
}
\label{fig:V1Ms}
\end{figure}

The effect of improving the momentum resolution is first investigated
while keeping the vertex resolution fixed at its baseline V1 value.
Figure~\ref{fig:V1Ms} compares the reconstructed mass distributions
obtained for V1M1, V1M2, and V1M4 using the selected reconstruction
strategy.

The fitted DCB core-width parameter decreases in central value from
$50.369\pm3.197~\mathrm{MeV}$ for V1M1 to
$49.847\pm2.748~\mathrm{MeV}$ for V1M2 and
$44.411\pm3.611~\mathrm{MeV}$ for V1M4. In contrast, the
model-independent measures remain nearly unchanged. The RMS values are
$223.201$, $223.458$, and $222.941~\mathrm{MeV}$, respectively,
while $W_{68}$ changes only from $376.018~\mathrm{MeV}$ to
$375.554~\mathrm{MeV}$ and $375.054~\mathrm{MeV}$. The corresponding numbers of retained events are $58{,}950$, $59{,}365$,
and $59{,}083$ for V1M1, V1M2, and V1M4, respectively, corresponding
to partial-reconstruction efficiencies of $54.07\%$, $54.46\%$, and
$54.20\%$. Thus, although the fitted core-width central value decreases
toward M4, the global spread and retained event fraction show little
variation across the V1 series.

\begin{figure}[htbp]
\centering
\includegraphics[width=0.99\textwidth]{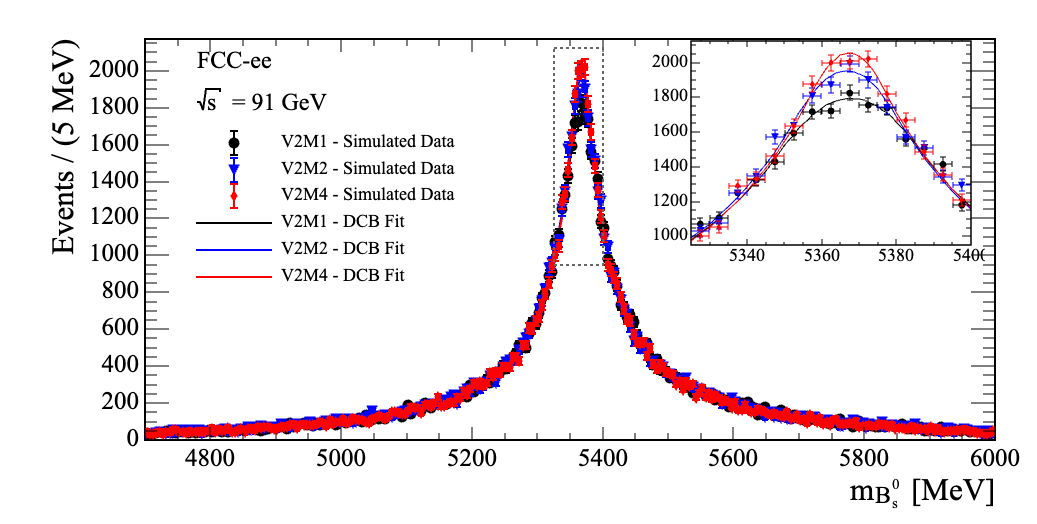}
\caption{
Reconstructed $B_s^0$ invariant-mass distributions and corresponding
DCB fits for the V2M1, V2M2, and V2M4 configurations using the
selected reconstruction strategy.
}
\label{fig:V2Ms}
\end{figure}

The vertex resolution is next improved by a factor of two to V2, while
the momentum resolution is again varied through the M1, M2, and M4
configurations. Figure~\ref{fig:V2Ms} compares the corresponding
reconstructed mass distributions.

At fixed M1 momentum resolution, improving the vertex resolution from
V1 to V2 changes both the fitted central component and the global
reconstruction measures. The fitted DCB core width decreases from
$50.369~\mathrm{MeV}$ for V1M1 to $30.831~\mathrm{MeV}$ for V2M1.
Over the same comparison, the RMS decreases from $223.201$ to
$203.432~\mathrm{MeV}$ and $W_{68}$ decreases from $376.018$ to
$304.325~\mathrm{MeV}$. The partial-reconstruction efficiency
simultaneously increases from $54.07\%$ to $64.66\%$. The
vertex-resolution improvement therefore affects not only the fitted
central component but also the broader reconstructed mass distribution
and the fraction of events retained by the reconstruction.

Within the V2 series, the fitted DCB core-width parameter decreases in
central value from $30.831\pm1.500~\mathrm{MeV}$ for V2M1 to
$24.720\pm1.617~\mathrm{MeV}$ for V2M2 and
$20.416\pm1.460~\mathrm{MeV}$ for V2M4. The model-independent widths
change much less. The RMS remains near $202$--$203~\mathrm{MeV}$,
while $W_{68}$ decreases from $304.325~\mathrm{MeV}$ for V2M1 to
$299.483~\mathrm{MeV}$ for V2M4. The partial-reconstruction efficiencies are $64.66\%$, $64.99\%$, and
$65.01\%$, corresponding to $70{,}491$, $70{,}848$, and $70{,}870$
retained events for V2M1, V2M2, and V2M4, respectively.

\begin{figure}[htbp]
\centering
\includegraphics[width=0.99\textwidth]{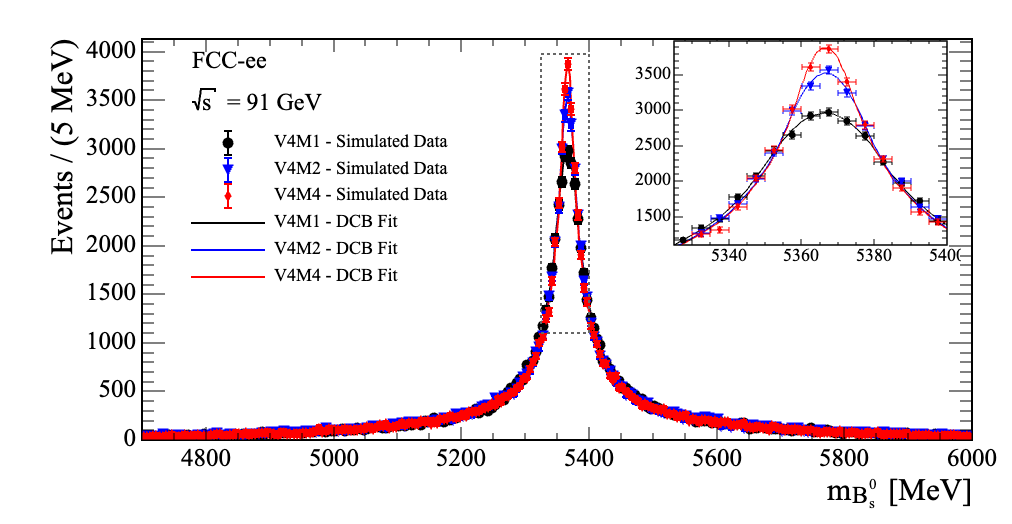}
\caption{
Reconstructed $B_s^0$ invariant-mass distributions and corresponding
DCB fits for the V4M1, V4M2, and V4M4 configurations using the
selected reconstruction strategy.
}
\label{fig:V4Ms}
\end{figure}

The vertex resolution is then improved further to V4, with the same
M1, M2, and M4 momentum-resolution configurations.
Figure~\ref{fig:V4Ms} shows the corresponding reconstructed mass
distributions.

At fixed M1 momentum resolution, improving the vertex resolution from
V2 to V4 again changes both the fitted central component and the global
reconstruction measures. The fitted DCB core width decreases from
$30.831~\mathrm{MeV}$ for V2M1 to $21.926~\mathrm{MeV}$ for V4M1.
The RMS decreases from $203.432$ to $189.046~\mathrm{MeV}$ and
$W_{68}$ decreases from $304.325$ to $244.497~\mathrm{MeV}$, while
the partial-reconstruction efficiency increases from $64.66\%$ to
$72.52\%$.

Within the V4 series, the fitted DCB core-width parameter decreases in
central value from $21.926\pm0.581~\mathrm{MeV}$ for V4M1 to
$15.620\pm0.584~\mathrm{MeV}$ for V4M2 and
$11.998\pm0.611~\mathrm{MeV}$ for V4M4. The model-independent
quantities again vary much less. The RMS decreases only from
$189.046$ to $187.571~\mathrm{MeV}$, while $W_{68}$ changes from
$244.497$ to $242.233~\mathrm{MeV}$. The partial-reconstruction efficiency increases modestly from $72.52\%$
($79{,}058$ events) to $73.30\%$ ($79{,}910$ events).

The post-fit Poisson-deviance results provide a separate assessment of
the empirical DCB description. The dependence on histogram scale
becomes particularly visible for the V4 configurations. For V4M2,
$D/\mathrm{ndf}$ changes from $1.117$ at a bin width of
$0.5~\mathrm{MeV}$ to $1.392$ at $5~\mathrm{MeV}$ and $1.973$ at
$10~\mathrm{MeV}$. For V4M4, the corresponding values are $1.089$,
$1.510$, and $2.199$. Thus, although the unbinned fits remain
numerically stable, the empirical DCB does not reproduce all features
of the reconstructed spectra equally well at every histogram scale.
The DCB is therefore treated as an empirical parameterization of the
reconstructed response rather than as an exact statistical model of
the full distribution.

The nominal detector-level DCB fits converge with minimization status
equal to zero and covariance quality equal to three, with small EDM
values. The HESSE covariance evaluation is included in the fitting
procedure. These quantities diagnose numerical convergence and the
quality of the fitted covariance matrices; they are distinct from the
post-fit goodness-of-fit assessment.

The numerical results for all studied configurations are summarized in
Tables~\ref{tab:dcb_fit_parameters},
\ref{tab:rms_w68_eff}, and
\ref{tab:poisson_deviance}.
Table~\ref{tab:dcb_fit_parameters} presents the fitted DCB parameters,
while Table~\ref{tab:rms_w68_eff} summarizes the corresponding RMS,
$W_{68}$, and partial-reconstruction efficiency.
The post-fit Poisson-deviance diagnostics, including $D$, the number of
degrees of freedom, and $D/\mathrm{ndf}$ for the tested histogram bin
widths, are reported separately in
Table~\ref{tab:poisson_deviance}.
The MC-truth results are included as reference benchmarks for comparison
with the detector-level configurations.

The fitted DCB core width decreases with improving momentum resolution, particularly in the V2 and V4 series. Since the detector configurations are based on the same generated events, their fitted widths are statistically correlated. Their fitted values of $\sigma$ are therefore
statistically correlated. The uncertainty on differences in $\sigma$
is evaluated using a paired event-level nonparametric
bootstrap~\cite{efron1979,davison1997}. The generated event is taken
as the resampling unit, and the same bootstrap multiplicity is applied
to that event across all detector configurations, thereby preserving
the event-by-event correspondence.

\newcommand{\fitval}[2]{%
    \makecell[c]{#1\\$\pm #2$}%
}

\begin{table}[!ht]
\centering
\caption{Best-fit double-sided Crystal Ball (DCB) parameters obtained from
the unbinned maximum-likelihood fits to the reconstructed $B_s^0$
invariant-mass distributions over the range $4700$--$6000~\mathrm{MeV}$.
The MC and V1M1 Cheated/Random rows correspond to the indicated reconstruction
strategies, while the detector configurations V1M1--V4M4 correspond to the
selected reconstruction strategy adopted for the detector-resolution study.
The quoted uncertainties are the fit uncertainties returned by
\textsc{RooFit}.}
\label{tab:dcb_fit_parameters}

\setlength{\tabcolsep}{4.0pt}
\renewcommand{\arraystretch}{1.05}

\begin{tabular}{|l|c|c|c|c|c|c|}
\hline
\textbf{Config.}
& $\boldsymbol{\mu}$ \textbf{(MeV)}
& $\boldsymbol{\sigma}$ \textbf{(MeV)}
& $\boldsymbol{\alpha_L}$
& $\boldsymbol{n_L}$
& $\boldsymbol{\alpha_R}$
& $\boldsymbol{n_R}$ \\
\hline

\makecell[l]{MC\\Cheated}
& \fitval{5366.770}{0.001}
& \fitval{0.00132}{0.00045}
& \fitval{0.00862}{0.00294}
& \fitval{1.8682}{0.0076}
& \fitval{0.00868}{0.00296}
& \fitval{1.8583}{0.0076}
\\
\hline

\makecell[l]{MC\\Random}
& \fitval{5366.769}{0.001}
& \fitval{0.16417}{0.00344}
& \fitval{1.3682}{0.0109}
& \fitval{0.9009}{0.0030}
& \fitval{1.3927}{0.0111}
& \fitval{0.8841}{0.0031}
\\
\hline

\makecell[l]{MC\\Selected}
& \fitval{5366.768}{0.001}
& \fitval{0.15314}{0.00325}
& \fitval{1.3471}{0.0109}
& \fitval{0.9175}{0.0030}
& \fitval{1.3350}{0.0112}
& \fitval{0.9219}{0.0031}
\\
\hline

\makecell[l]{V1M1\\Cheated}
& \fitval{5367.163}{0.570}
& \fitval{47.78}{1.71}
& \fitval{0.696}{0.029}
& \fitval{3.595}{0.145}
& \fitval{0.692}{0.028}
& \fitval{3.171}{0.121}
\\
\hline

\makecell[l]{V1M1\\Random}
& \fitval{5366.702}{0.907}
& \fitval{52.37}{3.08}
& \fitval{0.575}{0.035}
& \fitval{2.034}{0.084}
& \fitval{0.564}{0.034}
& \fitval{1.805}{0.074}
\\
\hline

V1M1
& \fitval{5366.795}{0.897}
& \fitval{50.37}{3.20}
& \fitval{0.562}{0.036}
& \fitval{2.110}{0.084}
& \fitval{0.518}{0.034}
& \fitval{2.196}{0.097}
\\
\hline

V1M2
& \fitval{5369.672}{0.857}
& \fitval{49.85}{2.75}
& \fitval{0.578}{0.032}
& \fitval{2.002}{0.074}
& \fitval{0.557}{0.033}
& \fitval{1.962}{0.082}
\\
\hline

V1M4
& \fitval{5367.343}{0.874}
& \fitval{44.41}{3.61}
& \fitval{0.518}{0.041}
& \fitval{2.077}{0.080}
& \fitval{0.481}{0.039}
& \fitval{2.102}{0.087}
\\
\hline

V2M1
& \fitval{5367.806}{0.469}
& \fitval{30.83}{1.50}
& \fitval{0.614}{0.028}
& \fitval{1.673}{0.037}
& \fitval{0.603}{0.028}
& \fitval{1.616}{0.037}
\\
\hline

V2M2
& \fitval{5366.719}{0.432}
& \fitval{24.72}{1.62}
& \fitval{0.551}{0.032}
& \fitval{1.612}{0.032}
& \fitval{0.532}{0.032}
& \fitval{1.608}{0.034}
\\
\hline

V2M4
& \fitval{5367.326}{0.401}
& \fitval{20.42}{1.46}
& \fitval{0.488}{0.031}
& \fitval{1.586}{0.029}
& \fitval{0.470}{0.030}
& \fitval{1.591}{0.030}
\\
\hline

V4M1
& \fitval{5366.555}{0.254}
& \fitval{21.93}{0.58}
& \fitval{0.793}{0.019}
& \fitval{1.344}{0.020}
& \fitval{0.746}{0.018}
& \fitval{1.381}{0.021}
\\
\hline

V4M2
& \fitval{5366.799}{0.212}
& \fitval{15.62}{0.58}
& \fitval{0.707}{0.022}
& \fitval{1.300}{0.016}
& \fitval{0.685}{0.022}
& \fitval{1.306}{0.017}
\\
\hline

V4M4
& \fitval{5366.588}{0.192}
& \fitval{12.00}{0.61}
& \fitval{0.622}{0.025}
& \fitval{1.287}{0.015}
& \fitval{0.591}{0.025}
& \fitval{1.310}{0.016}
\\
\hline

\end{tabular}

\end{table}

\newpage

\begin{table}[!ht]
\centering
\caption{Root-mean-square (RMS), central 68\% width, $W_{68}$, and partial
reconstruction efficiency, $\epsilon_{\rm PR}$, of the reconstructed
$B_s^0$ invariant-mass distributions within the
$4700$--$6000~\mathrm{MeV}$ analysis window. Here,
$W_{68}=q_{84}-q_{16}$, where $q_{16}$ and $q_{84}$ are the 16th and
84th percentiles of the reconstructed mass distribution. The partial
reconstruction efficiency is defined relative to the $109{,}016$ valid
generated events after removal of events affected by the generation
replication issue. The MC and V1M1 Cheated/Random rows correspond to
the indicated reconstruction strategies, while configurations
V1M1--V4M4 correspond to the selected reconstruction strategy adopted
for the detector-resolution study.}
\label{tab:rms_w68_eff}

\setlength{\tabcolsep}{10pt}
\renewcommand{\arraystretch}{1.05}

\newcommand{\configtwo}[2]{%
    \makecell[l]{#1\\#2}%
}

\newcommand{\configone}[1]{%
    \makecell[l]{#1\\\vphantom{Selected}}%
}

\begin{tabular}{|l|c|c|c|}
\hline
\textbf{Configuration}
& \textbf{RMS (MeV)}
& $\boldsymbol{W_{68}}$ \textbf{(MeV)}
& $\boldsymbol{\epsilon_{\rm PR}}$ \textbf{(\%)} \\
\hline

\configtwo{MC}{Cheated}
& 5.125
& 1.781
& 97.07 \\
\hline

\configtwo{MC}{Random}
& 168.638
& 128.906
& 90.62 \\
\hline

\configtwo{MC}{Selected}
& 160.923
& 106.614
& 92.27 \\
\hline

\configtwo{V1M1}{Cheated}
& 164.314
& 234.139
& 57.51 \\
\hline

\configtwo{V1M1}{Random}
& 228.853
& 391.826
& 52.77 \\
\hline

\configone{V1M1}
& 223.201
& 376.018
& 54.07 \\
\hline

\configone{V1M2}
& 223.458
& 375.554
& 54.46 \\
\hline

\configone{V1M4}
& 222.941
& 375.054
& 54.20 \\
\hline

\configone{V2M1}
& 203.432
& 304.325
& 64.66 \\
\hline

\configone{V2M2}
& 202.202
& 301.023
& 64.99 \\
\hline

\configone{V2M4}
& 202.246
& 299.483
& 65.01 \\
\hline

\configone{V4M1}
& 189.046
& 244.497
& 72.52 \\
\hline

\configone{V4M2}
& 188.569
& 243.110
& 73.05 \\
\hline

\configone{V4M4}
& 187.571
& 242.233
& 73.30 \\
\hline

\end{tabular}
\end{table}

\newpage

\begin{table}[H]
\centering
\captionsetup{font=small}
\caption{
Post-fit Poisson-deviance goodness-of-fit diagnostics for the DCB mass
distributions. V1M1--V4M4 correspond to the selected reconstruction
strategy unless otherwise indicated.
}
\label{tab:poisson_deviance}

\setlength{\tabcolsep}{8pt}
\renewcommand{\arraystretch}{1.10}

\begin{tabular}{|l|c|c|c|c|}
\hline
\textbf{Configuration}
& $\boldsymbol{\Delta m}$ \textbf{(MeV)}
& $\boldsymbol{D}$
& \textbf{ndf}
& $\boldsymbol{D/\mathrm{ndf}}$ \\
\hline

MC Cheated
&
\makecell[c]{0.125\\0.250\\0.500}
&
\makecell[c]{4862.527\\3712.076\\3136.485}
&
\makecell[c]{10393\\5193\\2593}
&
\makecell[c]{0.468\\0.715\\1.210}
\\
\hline

MC Random
&
\makecell[c]{0.125\\0.250\\0.500}
&
\makecell[c]{17194.038\\10716.503\\7659.812}
&
\makecell[c]{10393\\5193\\2593}
&
\makecell[c]{1.654\\2.064\\2.954}
\\
\hline

MC Selected
&
\makecell[c]{0.125\\0.250\\0.500}
&
\makecell[c]{17445.839\\11232.060\\8156.719}
&
\makecell[c]{10393\\5193\\2593}
&
\makecell[c]{1.679\\2.163\\3.146}
\\
\hline

V1M1 Cheated
&
\makecell[c]{0.500\\5\\10}
&
\makecell[c]{2876.377\\298.133\\147.024}
&
\makecell[c]{2593\\253\\123}
&
\makecell[c]{1.109\\1.178\\1.195}
\\
\hline

V1M1 Random
&
\makecell[c]{0.500\\5\\10}
&
\makecell[c]{2643.595\\272.722\\129.762}
&
\makecell[c]{2593\\253\\123}
&
\makecell[c]{1.020\\1.078\\1.055}
\\
\hline

V1M1
&
\makecell[c]{0.500\\5\\10}
&
\makecell[c]{2551.354\\273.829\\121.609}
&
\makecell[c]{2593\\253\\123}
&
\makecell[c]{0.984\\1.082\\0.989}
\\
\hline

V1M2
&
\makecell[c]{0.500\\5\\10}
&
\makecell[c]{2608.370\\256.939\\151.182}
&
\makecell[c]{2593\\253\\123}
&
\makecell[c]{1.006\\1.016\\1.229}
\\
\hline

V1M4
&
\makecell[c]{0.500\\5\\10}
&
\makecell[c]{2645.561\\266.493\\130.905}
&
\makecell[c]{2593\\253\\123}
&
\makecell[c]{1.020\\1.053\\1.064}
\\
\hline

V2M1
&
\makecell[c]{0.500\\5\\10}
&
\makecell[c]{2594.301\\262.445\\121.913}
&
\makecell[c]{2593\\253\\123}
&
\makecell[c]{1.001\\1.037\\0.991}
\\
\hline

V2M2
&
\makecell[c]{0.500\\5\\10}
&
\makecell[c]{2638.146\\319.109\\178.491}
&
\makecell[c]{2593\\253\\123}
&
\makecell[c]{1.017\\1.261\\1.451}
\\
\hline

V2M4
&
\makecell[c]{0.500\\5\\10}
&
\makecell[c]{2821.076\\287.301\\145.652}
&
\makecell[c]{2593\\253\\123}
&
\makecell[c]{1.088\\1.136\\1.184}
\\
\hline

V4M1
&
\makecell[c]{0.500\\5\\10}
&
\makecell[c]{2615.100\\300.967\\178.697}
&
\makecell[c]{2593\\253\\123}
&
\makecell[c]{1.009\\1.190\\1.453}
\\
\hline

V4M2
&
\makecell[c]{0.500\\5\\10}
&
\makecell[c]{2897.418\\352.290\\242.708}
&
\makecell[c]{2593\\253\\123}
&
\makecell[c]{1.117\\1.392\\1.973}
\\
\hline

V4M4
&
\makecell[c]{0.500\\5\\10}
&
\makecell[c]{2824.922\\381.919\\270.437}
&
\makecell[c]{2593\\253\\123}
&
\makecell[c]{1.089\\1.510\\2.199}
\\
\hline

\end{tabular}

\end{table}

\newpage

Each detector configuration retains one record for every original
generated event, identified by the same event number across
configurations. A successfully reconstructed event with a mass inside
the $4700$--$6000~\mathrm{MeV}$ analysis interval stores its
reconstructed mass, whereas an event that fails reconstruction or lies
outside this interval contributes no reconstructed mass. The event
records were checked to ensure that the event numbers are aligned
across all detector configurations.

For each bootstrap replica, $109,016$ event numbers are drawn with
replacement from the original set of $109,016$ events. An event can
therefore be selected more than once, while another event may not be
selected at all. Equivalently, each original event is assigned a
bootstrap multiplicity equal to the number of times it is drawn. The
same set of event multiplicities is used for every detector
configuration.

For a given configuration, the reconstructed mass of an accepted event
is entered into the unbinned DCB dataset according to this
multiplicity. For example, if an event is selected twice in a bootstrap
replica, its reconstructed mass enters the fit twice in a configuration
in which that event is accepted. If the same event fails reconstruction
or lies outside the mass window in another configuration, it
contributes no mass to that configuration. The number of reconstructed
masses entering an individual bootstrap fit can therefore differ among
the detector configurations even though the same resampling of the
original generated events is used.

The DCB fit is repeated for every configuration in each bootstrap
replica, yielding a new fitted value of $\sigma$. For each pair of
configurations, the difference
$\Delta\sigma=\sigma_A-\sigma_B$ is calculated within the same
replica. The procedure is repeated for 1000 bootstrap replicas, and the
standard deviation of the resulting distribution of
$\Delta\sigma$ is taken as the correlated bootstrap uncertainty,
$\delta_{\rm boot}(\Delta\sigma)$. The quantity
$Z_{\rm boot}=|\Delta\sigma|/\delta_{\rm boot}(\Delta\sigma)$
expresses the observed difference relative to this uncertainty. The
bootstrap therefore evaluates the statistical uncertainty of the
difference in fitted core widths while preserving the correlation
arising from their common generated events; it is not a
goodness-of-fit test of the DCB parameterization.

\begin{table}[!ht]
\centering
\caption{
Pairwise differences in the fitted DCB core-width parameter,
$\sigma$, for the momentum-resolution configurations within each
vertex-resolution series. The correlated uncertainty
$\delta_{\rm boot}$ is obtained from the standard deviation of the
paired-bootstrap distribution of $\Delta\sigma$ using 1000 bootstrap
replicas. The V1 comparisons contain 998 valid paired replicas, while
the V2 and V4 comparisons contain 1000.
}
\label{tab:bootstrap_sigma_differences}
\begin{tabular}{|c|c|c|c|c|}
\hline
\textbf{Series} &
\textbf{Comparison} &
$\boldsymbol{\Delta\sigma}$ \textbf{(MeV)} &
$\boldsymbol{\delta_{\rm boot}}$ \textbf{(MeV)} &
$\boldsymbol{Z_{\rm boot}}$ \\
\hline
V1 & M1--M2 & 0.522 & 5.476 & 0.095 \\
V1 & M1--M4 & 5.958 & 5.441 & 1.095 \\
V1 & M2--M4 & 5.436 & 5.527 & 0.984 \\
\hline
V2 & M1--M2 & 6.112 & 2.503 & 2.442 \\
V2 & M1--M4 & 10.416 & 2.159 & 4.825 \\
V2 & M2--M4 & 4.304 & 2.279 & 1.888 \\
\hline
V4 & M1--M2 & 6.306 & 0.786 & 8.022 \\
V4 & M1--M4 & 9.927 & 0.896 & 11.076 \\
V4 & M2--M4 & 3.622 & 0.922 & 3.927 \\
\hline
\end{tabular}
\end{table}

The paired-bootstrap results determine which of the apparent changes in
the fitted core width are statistically resolved. Within the V1
series, none of the pairwise differences is resolved. In particular,
the V1M1--V1M4 difference is
$5.958\pm5.441~\mathrm{MeV}$, corresponding to
$Z_{\rm boot}=1.095$. The decrease of the fitted core-width central
value within V1 is therefore not statistically established by the
present sample.

Within the V2 series, the V2M1--V2M4 difference is
$10.416\pm2.159~\mathrm{MeV}$, corresponding to
$Z_{\rm boot}=4.825$. The M1--M2 and M2--M4 comparisons give
$Z_{\rm boot}=2.442$ and $1.888$, respectively. The endpoint
difference between M1 and M4 is therefore clearly resolved, while the
individual intermediate changes provide weaker separation.

The momentum-resolution dependence of the fitted core width is most
clearly resolved within the V4 series. The V4M1--V4M2,
V4M1--V4M4, and V4M2--V4M4 comparisons give
$Z_{\rm boot}=8.022$, $11.076$, and $3.927$, respectively. The
decrease of the fitted DCB core-width parameter with improved momentum
resolution is therefore statistically resolved throughout the V4
series.

Taken together, the results indicate different sensitivities to vertex-
and momentum-resolution improvements. Improving the vertex resolution
at fixed M1 produces substantial changes not only in the fitted DCB
core-width parameter but also in the RMS, $W_{68}$, and
partial-reconstruction efficiency. From V1M1 to V2M1 and V4M1, for
example, the RMS decreases from $223.201~\mathrm{MeV}$ to
$203.432~\mathrm{MeV}$ and $189.046~\mathrm{MeV}$, while $W_{68}$
decreases from $376.018~\mathrm{MeV}$ to $304.325~\mathrm{MeV}$ and
$244.497~\mathrm{MeV}$. The corresponding efficiencies increase from
$54.07\%$ to $64.66\%$ and $72.52\%$.

The effect of improving the momentum resolution at fixed vertex
resolution is more localized. No statistically resolved change in the
fitted DCB core width is observed within the V1 series, whereas the
M1--M4 differences are clearly resolved for V2 and V4. Even in these
series, however, the corresponding changes in RMS, $W_{68}$, and
partial-reconstruction efficiency remain comparatively small.
Consequently, the vertex resolution has the stronger influence on the
global reconstruction performance, whereas improvements in momentum
resolution are manifested primarily through the fitted central DCB
component. The latter should not be interpreted as an equivalent
narrowing of the complete reconstructed mass distribution.

The comparison with the MC-truth study further indicates that the
overall spread of the reconstructed mass distribution is not determined
by detector resolution alone, but also reflects the quadratic ambiguity
and the kinematic structure inherent to the partial-reconstruction
procedure.


\section{Summary and Conclusions}

The partial reconstruction of
$B_s^0\rightarrow\phi(K^+K^-)\eta'(\pi^+\pi^-\eta)$ was studied in an
FCC-ee environment without explicitly reconstructing the neutral $\eta$.
The missing-particle kinematics are recovered from the topological and
kinematic constraints of the decay, resulting in a quadratic ambiguity.
Among the experimentally applicable prescriptions considered at MC-truth
level, selecting the solution whose reconstructed $B_s^0$ energy is closest
to the nominal beam energy gives better overall reconstruction performance
than random solution selection. The selected strategy retains $100{,}590$
events compared with $98{,}792$ for the randomized treatment and gives
smaller RMS and $W_{68}$ values. It is therefore adopted for the
detector-resolution study.

The detector-level results show that improvements in vertex resolution
have the strongest effect on the global reconstructed mass response.
At fixed M1 momentum resolution, improving the vertex configuration from
V1M1 to V4M1 reduces the RMS from $223.201$ to $189.046~\mathrm{MeV}$
and $W_{68}$ from $376.018$ to $244.497~\mathrm{MeV}$, while the
partial-reconstruction efficiency increases from $54.07\%$ to $72.52\%$.
Thus, improved vertex resolution affects both the overall spread of the
reconstructed mass distribution and the fraction of events successfully
retained.

The effect of improving the momentum resolution is more localized.
Within the V1 series, the paired-bootstrap analysis finds no statistically
resolved change in the fitted DCB core-width parameter. In contrast, the
M1--M4 differences are clearly resolved for V2 and V4, with
$Z_{\rm boot}=4.825$ and $11.076$, respectively. The corresponding
changes in the model-independent RMS, $W_{68}$, and reconstruction
efficiency remain comparatively small. For example, within the V4 series,
the RMS changes from $189.046$ to $187.571~\mathrm{MeV}$ and $W_{68}$
from $244.497$ to $242.233~\mathrm{MeV}$ as the momentum resolution is
improved from M1 to M4. Momentum-resolution improvements therefore
primarily modify the fitted central DCB component and should not be
interpreted as an equivalent narrowing of the complete reconstructed
mass distribution.

The DCB fits provide a useful empirical characterization of the
reconstructed mass response, although the post-fit Poisson-deviance
diagnostics show that the level of agreement depends on the histogram
scale, particularly for the higher-resolution configurations. The global
mass response is therefore characterized using the fitted DCB parameters
together with the model-independent RMS, $W_{68}$, and
partial-reconstruction efficiency. Overall, the study demonstrates that
partial reconstruction can recover a substantial fraction of
$B_s^0\rightarrow\phi\eta'$ candidates without explicit reconstruction
of the neutral $\eta$. Within the detector scenarios considered here,
vertex resolution has the stronger influence on the overall reconstruction
performance, while improved momentum resolution provides additional
localization of the central reconstructed mass component.

\section*{Acknowledgments}

The authors gratefully acknowledge the Office of Research of
Mindanao State University--Iligan Institute of Technology
for its support and provision of resources that contributed
to the conduct of this research. The lead author also expresses
his appreciation to the University of the Philippines Mindanao
for providing a supportive academic environment that enables
faculty members to pursue advanced studies and research.

\normalsize
\bibliography{main}


\end{document}